\documentclass[english,aps,prd,epsfig]{revtex4}
\usepackage[T1]{fontenc}
\usepackage[latin1]{inputenc}
\usepackage{babel}
\usepackage{graphics}
\makeatletter

\providecommand{\LyX}{L\kern-.1667em\lower.25em\hbox{Y}\kern-.125emX\@}
\let\SF@@footnote\footnote
\def\footnote{\ifx\protect\@typeset@protect
    \expandafter\SF@@footnote
  \else
    \expandafter\SF@gobble@opt
  \fi
}
\expandafter\def\csname SF@gobble@opt \endcsname{\@ifnextchar[
  \SF@gobble@twobracket
  \@gobble
}
\edef\SF@gobble@opt{\noexpand\protect
  \expandafter\noexpand\csname SF@gobble@opt \endcsname}
\def\SF@gobble@twobracket[#1]#2{}

\newcommand{\bee}{\begin{equation}}
\newcommand{\ee}{\end{equation}}
\newcommand{\beea}{\begin{eqnarray}}
\newcommand{\eea}{\end{eqnarray}}

\makeatother
\begin{document}

{\centering\textbf{\Large Improving the Chiral Properties of Lattice Fermions} {\Large }\Large \par}

\author{Thomas DeGrand, Anna Hasenfratz}

\address{Department of Physics, University of Colorado, Boulder, CO 80309
USA }

\author{Tam\'{a}s G.\ Kov\'{a}cs}

\address{Department of Theoretical Physics,
     University of P\'ecs,
     H-7624 P\'ecs, Ifj\'us\'ag \'utja 6, Hungary}

\date{\today{}}

\begin{abstract}
The chiral properties of lattice fermions can be improved by
altering either their fermion-gauge coupling or the pure gauge
part of the action (or both).
 Using both perturbation theory and nonperturbative 
simulation, we compare a simple alteration of the gauge action
(which encompasses the Wilson, Symanzik,
Iwasaki, and DBW2 actions), and HYP-blocked links in the fermion action.
Perturbative tests include calculations of the potential, flavor-changing
quark
scattering amplitudes, and matching factors for currents.
Non-perturbative tests include the potential,
 measurements of flavor symmetry breaking for
staggered fermions, the behavior of topological objects,
 and properties of overlap actions.
Our results display
 the bad properties of these actions as well as their good ones.
\end{abstract}
\maketitle

\section{Introduction}

The chiral symmetry breaking of lattice fermions is the most serious
lattice artifact we face in numerical simulations. Most dynamical
simulations use staggered or Wilson-like fermions, where chiral symmetry
breaking is explicit and difficult to control. In the case of staggered
fermions lattice artifacts
 induce flavor symmetry breaking, resulting in non-Goldstone
pions that can be significantly heavier than the Goldstone particle.
In the case of Wilson-like fermions the effect of chiral symmetry
breaking is most observable in the occurence of exceptional configurations.
Domain wall fermions, which in theory are chiral, have a residual
chiral symmetry breaking due to the finiteness of the fifth lattice
direction. Overlap fermions are chiral, no matter what non-chiral
action they are constructed from. However, the computational requirements
needed
to evaluate the overlap operator can be significantly lowered if it
is based on a near-chiral action.

A systematic improvement program modifies the pure gauge part of the
action, the fermionic action, and the gauge connections of the fermions,
to construct a chirally improved action with reduced lattice artifacts.
Less ambitious programs might change only part of the action at a
time, according to the improvement needs of the program. 

Several possibilities have been explored to improve the chiral symmetry
of fermionic actions. They are all based on the observation that most
of the chiral violation come from the short scale, plaquette level
vacuum fluctuations of the gauge field. One option is to smear the
gauge connection of the fermions, reducing the coupling of
fluctuations to the fermions. This option
has been used  with staggered
 fermions \cite{Toussaint:2001zc, Hasenfratz:2002jn}.
The two smearing transformations that are used most extensively are
the Fat7/Asqtad \cite{Lepage:1998vj, Orginos:1998ue, Orginos:1999cr}
and HYP smearings \cite{Hasenfratz:2001hp}. Both of these transformations
are \( O(a^{2}) \) perturbatively improved, though only the Asqtad smearing
was designed perturbatively. The HYP smearing is optimized non-perturbatively. 

The other option that has been studied lately is to modify the pure
gauge action such a way that the creation of small scale vacuum fluctuations
is dynamically suppressed. The Iwasaki action \cite{Iwasaki:1983ck}
has been used extensively by CP-PACS both in dynamical Wilson and
domain wall  fermion simulations  \cite{Iwa:AliKhan:1999ib, DW:AliKhan:2000iv}
while the DBW2 action \cite{DBW2:Takaishi:1996xj, DBW2:deForcrand:1999bi}
was chosen by the Columbia-BNL group in their
domain wall fermion simulations \cite{DW-DBW2:Orginos:2001xa}. The
Iwasaki and especially the DBW2 actions have larger perturbative lattice
corrections than the Wilson plaquette action though in numerical tests
that did not seem to increase the lattice artifacts. 

The advantage for numerical simulations
 of modifying the gauge action instead of smearing the
gauge connection is obvious. Almost any gauge
action can be simulated faster than a complicated fermionic action,
though with the recently developed partial-global stochastic Metropolis
(PGSM) update even projected smeared link fermions can be simulated
effectively \cite{Hasenfratz:2002jn, Hasenfratz:2002ym, Alexandru:2002jr}.
The important  question is if the improvement offered by the modified gauge actions
is sufficient, and even more, if, in addition to the improved
chiral symmetry any unwanted lattice artifacts are introduced by these
actions.

Our goal in this paper is to compare different physical properties,
 both perturbative
and non-perturbative,  of the smeared and modified gauge
actions, and possibly to predict which choice is going to give the
most efficient approach of the continuum limit.

In Sec.~2 we introduce the actions we study. A perturbative
analysis of the properties of these actions is carried out in Sec.~3.
In Sec.~4 we show simulation results for the heavy quark potential.
The topological properties of these actions are discussed in Sec.~5.
Flavor symmetry violations for staggered actions are shown in Sec.~6.
Sec.~7 discusses some properties of overlap actions built using HYP
links or in the background of gauge fields with the gauge actions we study.
Our conclusions are found in Sec.~8.

\section{The Actions}

In the following we consider a family of actions consisting of the
\( 1\times 1 \) plaquette and the \( 1\times 2 \) planar loop \begin{equation}
\label{eq:actions}
S_{g}(U)=\frac{\beta }{3}[c_{0}\sum _{n,\mu <\nu }W_{\mu \nu }^{1\times 1}(n)+c_{1}\sum _{n,\mu \neq \nu }W_{\mu \nu }^{1\times 2}(n)]
\end{equation}
 with the normalization condition \( c_{0}+8c_{1}=1 \). The coefficient
\( c_{1} \) can vary,  giving the different
actions\begin{eqnarray}
c_{1}= & 0 & \qquad \rm {Wilson}\\
 & -1/12 & \qquad \rm {tree\: level\: Symanzik}\\
 & -0.331 & \qquad \rm {Iwasaki}\label{eq:c1coeff} \\
 & -1.4088 & \qquad \rm {DBW2}.
\end{eqnarray}
 The tree level Symanzik action  is \( O(a^{4}) \)
improved \cite{Weisz:1983zw}, while both the Iwasaki and DBW2 actions
have opposite \( O(a^{2}) \) corrections than the Wilson plaquette
action. Some of the scaling properties of the Iwasaki action have
been studied in \cite{Iwa:AliKhan:1999ib} while the DBW2 action was
investigated in \cite{DBW2:Takaishi:1996xj, DW-DBW2:Orginos:2001xa}.
In our numerical symulations we  use
 the 1-loop tadpole improved Symanzik action
\cite{LW:Luscher:1985xn, LW-tadpole:Bliss:1996wy}\begin{equation}
\label{eq:sym1l}
S_{\rm {Sym1l}}(U)=\frac{\beta_{1\times 1} }{3}\sum _{n,\mu <\nu }W_{\mu \nu }^{1\times 1}(n)+
\frac{\beta_{1\times 2} }{3}\sum _{n,\mu \neq \nu }W_{\mu \nu }^{1\times 2}(n)+
\frac{\beta_{pg} }{3}\sum _{n,\mu \ne \nu \ne \rho }W_{\mu \nu \rho }^{pg}
\end{equation}
 where \( W^{pg} \) is a six-link parallelogram with links running
around the opposing edges of the cube. The \( \beta \) coefficients
for \( S_{\rm {Sym1l}} \) are tadpole improved according to the
plaquette expectation value. 

In addition to the different gauge actions we will consider two smearing
transformations for the fermion-gauge field interaction. 
The Fat7 smearing replaces the usual one-link coupling with
 a linear combination
of gauge loops up to length seven. It has
 perturbatively determined coefficients
which remove the flavor changing gluons at the edges of the Brillouin
zone for staggered fermions. When an additional 5-link term is added
to the Fat7 smearing the resulting action is \( O(a^{2}) \) improved
in the fermion-gluon connection \cite{Lepage:1998vj}. With tadpole
boosted coefficients this leads to the Asqtad smearing transformation
\cite{Orginos:1999cr}. (One should note that the Asqtad staggered
action also has a third nearest neighbor Naik term, which we do not
include here.) The Asqtad smeared links are not unitary -- they are
simply the linear combinations of the extended gauge paths. 

Our second smearing transformation is HYP smearing \cite{Hasenfratz:2001hp}.
HYP smeared links are constructed from three levels of modified, SU(3)
projected APE blocking steps in a way that makes the transformation
local and smooth. While the HYP smearing is non-perturbatively optimized,
its coefficients can just
as well be tuned to give perturbative improvement.

\section{Perturbative Considerations}

\subsection{Preliminaries}

The gauge actions we study have propagators which obey the equation
\begin{equation}
\label{eq:OFM}
[{1\over {\xi -1}}\hat{k}_{\mu }\hat{k}\nu +\sum _{\rho }(\hat{k}_{\rho }\delta _{\mu \nu }-\hat{k}_{\mu }\delta _{\rho \mu }q_{\mu \rho }\hat{k}_{\rho })]G_{\mu \nu }=\delta _{\mu \nu }
\end{equation}
 where \begin{equation}
q_{\mu \nu }=(1-\delta _{\mu \nu })(1-c_{1}(\hat{k}_{\mu }^{2}+\hat{k}_{\nu }^{2}))
\end{equation}
 and \( \hat{k}_{\mu }=2\sin (k_{\mu }/2) \). \( \xi  \) is the
gauge-fixing term: \( \xi =1 \) is Feynman gauge. The Wilson gauge
action propagator is \begin{equation}
G_{\mu \nu }=[\delta _{\mu \nu }+(\xi -1){{\hat{k}_{\mu }\hat{k}_{\nu }}\over {\hat{k}^{2}}}]{1\over {\hat{k}^{2}}}.
\end{equation}
 For other actions we merely numerically invert Eq. \ref{eq:OFM}
to construct the propagator.

We will be concerned only with unitary fat links, gauge connections
which are themselves elements of the gauge group, even though they
may be built of sums of products of the original thin links of the
simulation. For smooth fields the fat links have an expansion \( V_{\mu }(x)=1+iaB_{\mu }(x)+\dots  \)
and the original thin links have an expansion \( U_{\mu }(x)=1+iaA_{\mu }(x)+\dots  \).
For computations of 2- and 4-quark operator renormalization/matching
constants at one loop, only the linear part of the relation between
fat and thin links is needed \cite{Bernard:1999kc}, and it can be
parameterized as \begin{equation}
\label{eq:linA}
B_{\mu }(x)=\sum _{y,\nu }h_{\mu \nu }(y)A_{\nu }(x+y)\, \, .
\end{equation}
 Quadratic terms in (\ref{eq:linA}), which would only be relevant
for tadpole graphs, appear as commutators and therefore do not contribute,
since tadpoles are symmetric in the two gluons. In momentum space,
the convolution of Eq.~(\ref{eq:linA}) becomes a form factor \begin{equation}
\label{eq:linkBA}
B_{\mu }(q)=\sum _{\nu }\tilde{h}_{\mu \nu }(q)A_{\nu }(q)\, \, .
\end{equation}
If all gluon lines start and end on fermion lines, then, effectively,
the gluon propagator changes into \begin{equation}
\label{eq:GEFF}
{\mathcal{G}}_{\mu \nu }\rightarrow \tilde{h}_{\mu \lambda }G_{\lambda \sigma }\tilde{h}_{\sigma \nu }.
\end{equation}
Obviously, this means that in perturbation theory, as far as the fermions are
concerned, fattening the links in the fermion action is equivalent
to altering the gauge action.

The family of smearings including HYP links \cite{Hasenfratz:2001hp}
and the order-\( a^{2} \) improved link \cite{Lepage:1998vj} which,
when augmented by tadpole improvement, gives the {}``Asqtad{}''
link used by the MILC collaboration~\cite{Orginos:1999cr}, have
\begin{equation}
\label{eq:impvert}
\tilde{h}_{\mu \nu }=\delta _{\nu ,\mu }D_{\mu }(k)+(1-\delta _{\nu ,\mu })G_{\nu ,\mu }(k)\, .
\end{equation}
 The diagonal and off-diagonal couplings can be decomposed, respectively,
as \begin{equation}
\label{eq:diag}
D_{\mu }(k)=1-{d_{1}\over 4}\sum _{\nu \ne \mu }{\hat{k}}_{\nu }^{2}+{d_{2}\over 16}\sum _{\nu <\rho \atop \nu ,\rho \ne \mu }{\hat{k}}_{\nu }^{2}{\hat{k}}_{\rho }^{2}-{d_{3}\over 64}{\hat{k}}_{\nu }^{2}{\hat{k}}_{\rho }^{2}{\hat{k}}_{\sigma }^{2}-{d_{4}\over 16}\sum _{\nu \ne \mu }{\hat{k}}_{\nu }^{4}\, ,
\end{equation}
 and \begin{eqnarray}
G_{\nu ,\mu }(k) & = &
\frac{ {\hat{k}}_{\mu }{\hat{k}}_{\nu }}{4}\widetilde{G}_{\nu ,\mu }(k)\\
\widetilde{G}_{\nu ,\mu }(k) & = & d_{1}-d_{2}\frac{({\hat{k}}_{\rho }^{2}+{\hat{k}}_{\sigma }^{2})}{8}+d_{3}\frac{{\hat{k}}_{\rho }^{2}{\hat{k}}_{\sigma }^{2}}{12}+d_{4}\frac{{\hat{k}}_{\nu }^{2}}{4}\, ,\label{eq:offdiag} 
\end{eqnarray}
 where all indices (\( \mu ,\nu ,\rho ,\sigma  \)) are different.

The coefficients \( d_{1-4} \) distinguish the different choices
of links: 

\begin{enumerate}
\item Fat-7 links: \begin{equation}
d_{1}=1,\quad d_{2}=1,\quad d_{3}=1,\quad d_{4}=0.
\end{equation}

\item \( O(a^{2}) \) improved links: \begin{equation}
d_{1}=0,\quad d_{2}=1,\quad d_{3}=1,\quad d_{4}=1.
\end{equation}

\item HYP smeared links: \begin{equation}
d_{1}=(2/3)\alpha _{1}(1+\alpha _{2}(1+\alpha _{3})),\quad d_{2}=(4/3)\alpha _{1}\alpha _{2}(1+2\alpha _{3}),\quad d_{3}=8\alpha _{1}\alpha _{2}\alpha _{3},\quad d_{4}=0.
\end{equation}
 There are two interesting choices for the \( \alpha _{i} \). The
first was determined in Ref.~\cite{Hasenfratz:2001hp} using a non-perturbative
optimization procedure: \( \alpha _{1}=0.75 \), \( \alpha _{2}=0.6 \)
\( \alpha _{3}=0.3 \). The second is chosen so to remove \( O(a^{2}) \)
flavor-symmetry breaking couplings at tree level. This gives the same \( d_i \)
as for Fat-7 links. 
\end{enumerate}

\subsection{Static Potential}

\begin{figure}
{\centering \resizebox*{10cm}{!}{\includegraphics{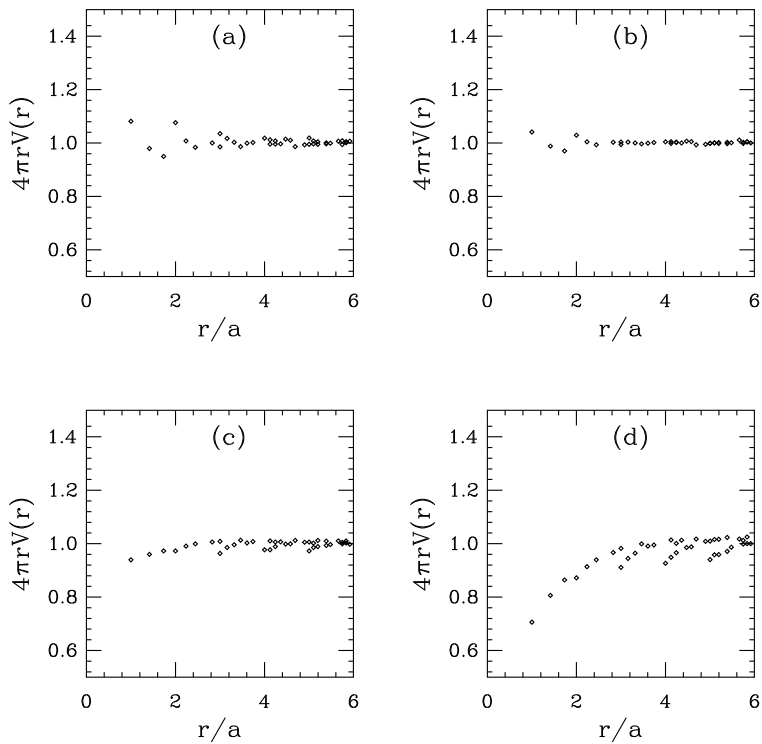}} \par}

\caption{The scaled static potential \protect\( 4\pi rV(r)\protect \) for
several lattice actions: (a) Wilson (\protect\( c_{1}=0\protect \)),
(b) tree level Symanzik (\protect\( c_{1}=-1/12\protect \)), (c)
Iwasaki (\protect\( c_{1}=-0.331\protect \))(d) DWB2 (\protect\( c_{1}=-1.4088\protect \)). }

\label{fig:vr}
\end{figure}

With the gluon propagator we can immediately compute the static
Coulomb  potential,
\begin{equation}
\label{eq:pert_pot}
V_c(r)=\int _{ak}\exp (i\vec{k}\cdot r)G_{00}(q_{\mu }=(0,\vec{k}))
\end{equation}
 where \( \int _{ak}=\prod _{j}\int _{-\pi }^{\pi }d(ak_{j})/(2\pi ) \)
will be the symbol for integration over the (rescaled) momentum hypercube.
With our conventions, the continuum potential is \( V(r)=1/(4\pi r) \),
and so plotting the rescaled lattice potential \( 4\pi rV(r) \) immediately
exposes the lattice artifacts of a particular action. We show results
for this quantity in Fig. \ref{fig:vr} for the four candidate actions
of our study. The tree level Symanzik action has no \( O(a^{2}) \)
nor \( O(a^{4}) \) discretization errors and also has the smallest
scaling violations. The other actions have \( O(a^{4}) \) scaling
violations, which should be of opposite sign for the Wilson action
versus the Iwasaki and DWB2 actions. This is seen in the figure. However,
the most noticable feature of the potential is the systematic distortion
of \( 4\pi rV(r) \) below unity at small lattice distance
as the coefficient
\( c_{1} \) becomes more negative. The results of this calculation
strongly disfavor use of a large negative value of \( c_{1} \)
for any lattice simulations with physics related to the short distance
part of the potential.

\begin{figure}
{\centering \resizebox*{10cm}{!}{\includegraphics{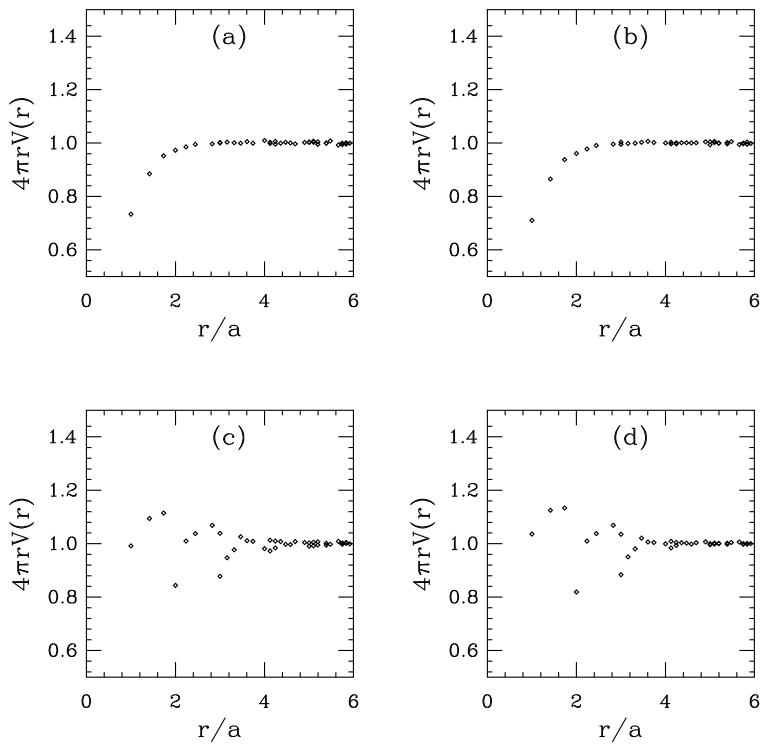}} \par}

\caption{The scaled static {}``smeared potential{}'' \protect\( 4\pi rV(r)\protect \)
for two lattice actions and two smearing functions: (a) Wilson (\protect\( c_{1}=0\protect \))
gauge action with HYP blocking, (b) tree level Symanzik (\protect\( c_{1}=-1/12\protect \))
gauge action with HYP blocking, (c) Wilson (\protect\( c_{1}=0\protect \))
gauge action with Asqtad blocking, (d) tree level Symanzik (\protect\( c_{1}=-1/12\protect \))
gauge action with Asqtad blocking, }

\label{fig:vrh}
\end{figure}

The potential itself is, of course, unaffected by any fattening of
the fermion's gauge connection, but we can also define a {}``smeared
potential,{}'' in which the gluon propagator \( G_{00} \) is replaced
by \( {\mathcal{G}}_{00} \). This quantity has the physical interpretation
of the potential seen by a heavy lattice quark whose gauge connection
is a fat link. Results for two gauge actions (Wilson and tree level
Symanzik) and two smearings, HYP and Asqtad, are shown in Fig. \ref{fig:vrh}.

Both of these smearings distort the lattice potential at small \( r/a \).
The immediate conclusion that one draws from these pictures is that
one should not do simulations involving heavy quarks with smeared
links -- the loss of Coulomb behavior at short distance will allow
the wave function of a heavy quark-antiquark bound state to spread
out, the value of the wave function at the origin will be small, and
hyperfine splittings will collapse. This effect is readily seen in
simulations.

A particular example of the danger of fattening heavy quarks is seen
in the recent MILC study of heavy quark-light quark decay constants,
\( f_{D} \) and \( f_{B} \) \cite{Bernard:2002pc}. One of the data
sets collected by these authors used fermions fattened with a large
amount of APE-smearing \cite{APE:Albanese:1987ds} (\( c=0.45 \),
\( N=10 \) in the conventions of Ref. \cite{Bernard:1999kc}). This
amount of smearing produces a noticable suppression of the static
potential out to \( r/a\simeq 3-4 \). The authors observed a twenty
per cent reduction in \( f_{B} \). This effect presumably
would go away at smaller
lattice spacing, but comparing Fig. \ref{fig:vrh}, one would need
to halve the lattice spacing to reduce the lattice artifact to the
level of HYP blocking.

However, the motivation for using smeared links in a fermion action
is, by and large, to improve the chiral properties of the action.
This is physics applicable to light quarks, not heavy ones. Many simulations
of light quarks with various degrees of fattening show no ill
effects on spectroscopy or on matrix elements -- generally, improvement
of scaling is observed \cite{Toussaint:2001zc,Hasenfratz:2002jn}.
 Fattening is something which can be done selectively,
in a mass-dependent way. Altering the gauge action will affect quarks
regardless of their mass.

Finally, we call the reader's attention to the rather large lattice
spacing artifacts of the Asqtad-smeared potential as compared to the
HYP- smeared potential (or any of the usual unsmeared potentials).

\subsection{One-loop Perturbative Matching Factors}

A matrix element of an observable computed using one regularization
(\( \overline{MS} \)) is related in perturbation theory to a linear
combination of observables computed using another (lattice) regularization
through a matrix of matching coefficients \( Z_{nm} \)\begin{equation}
\label{eq:ZFACTOR}
\langle f|O^{cont}_{n}(\mu )|i\rangle _{\overline{MS}}=a^{D}\sum _{m}Z(\mu ,a)_{nm}\langle f|O^{latt}(a)_{m}|i\rangle 
\end{equation}
 where each coefficient is a difference between (the finite part
of a) continuum-regulated
and a lattice-regulated  expression \begin{equation}
\label{eq:DEFINEZ}
Z_{nm}(\mu ,a,m)=1+{{g^{2}}\over {16\pi ^{2}}}(\Delta ^{F}_{\overline{MS}}-
\Delta _{latt}).
\end{equation}
 (\( g^{2} \) is the squared coupling). For currents, it is customary
to divide out the quadratic Casimir \( C_{F} \) and to present results
for \( z=(\Delta ^{F}_{\overline{MS}}-\Delta _{latt})/C_{F} \). For
four-quark operators (for weak-interaction matrix elements, for example),
it is customary just to write \( b=\Delta ^{F}_{\overline{MS}}-\Delta _{latt} \).
Generally, in the context of perturbation theory, one attempts to
design actions, operators, and methods of performing perturbative
calculations \cite{Lepage:1993xa} so that the \( z \) or \( b \)
coefficients are minimized. The reader might recall that at typical
lattice spacings studied today, typical choices for \( g^{2}/{4\pi } \)
range from 0.1-0.2, so a \( z \) or \( b \) of about 20 implies
a  twenty to forty per cent effect from a one loop calculation--perhaps
a bit large for comfort. Typically, with fat links one can reduce
these numbers by an order of magnitude, as seen by perturbative calculations\cite{Bernard:1999kc, LeeSharpe:Lee:2002bf, DeGrand:2002va}
or nonperturbative simulation \cite{DeGrand:1999gp}.

We have computed one loop perturbative matching factors for local
currents \( \bar{\psi }(x)\Gamma \psi (x) \)--for the vector (V),
axial vector (A), scalar (S) and pseudoscalar (P) currents, as well
as the matching factors for the four-fermion operators \( O_{\pm }=O_{1}\pm O_{2} \),
built from \begin{equation}
O=(\bar{q}^{(1)}_{\alpha }\Gamma _{1}q^{(2)}_{\beta })\otimes (\bar{q}^{(3)}_{\gamma }\Gamma _{2}\hat{q}^{(4)}_{\delta }).
\end{equation}
where \( \Gamma _{1}=\Gamma _{2}=\gamma _{\mu }(1-\gamma _{5}) \). \( O_{\pm }=O_{1}\pm O_{2} \); 
 if color labels \( \alpha =\delta  \), \( \beta =\gamma  \),
\( O=O_{1} \); if color labels \( \alpha =\beta  \), \( \gamma =\delta  \),
\( O=O_{2} \). (These operators have no penguin contributions.)

An important indicator of chiral improvement
 \cite{Gupta:1997yt} for nonchiral actions
is the difference \( z_{V}-z_{A} \) = \( (z_{P}-z_{S})/2 \): the
mixing of four-fermion operators into the opposite chirality sector
is controlled by this quantity.

Our results are shown in Tables \ref{tab:currcl}, \ref{tab:currnai}.
For operators with anomalous dimensions (all but the vector and axial
vector currents) our results are for the case (lattice spacing \( a\times  \)
regularization point \( \mu )=1 \). We see that as \( c_{1} \) becomes
increasingly negative, the \( z \) and \( b \) coefficients generally
shrink.

We next repeat these calculations, but now with fermions with HYP-smeared
gauge connections. Tables \ref{tab:currclh}, \ref{tab:currnaih}
show a dramatic reduction in the \( z \) and \( b \) coefficients,
even when the Wilson gauge action is used. With the reduction of these
numbers comes also a reduction in the difference \( z_{V}-z_{A} \)
for clover fermions.

From the point of perturbative theory for matching factors, the conclusion
is clear: it is much more efficient to fatten the fermion gauge connections
than to increase \( c_{1} \) in the gauge action.

\begin{table}
\begin{tabular}{|c|c|c|c|c|}
\hline 
&
 Wilson &
Tree level Sym&
 Iwasaki &
 DWB2 \\
\hline 
 \( Z_{V} \)&
 -15.33 &
 -11.91 &
 -7.44 &
 -3.03 \\
\hline
 \( Z_{A} \)&
 -13.79 &
 -10.72 &
 -6.71 &
 -2.75 \\
\hline
 \( Z_{S} \)&
 -19.31 &
 -15/08 &
 -8.90 &
 -1.04 \\
\hline 
 \( Z_{P} \)&
 -22.38 &
 -17.47 &
 -10.36 &
-1.59 \\
\hline
 \( Z_{+} \)&
 -36.63 &
 -28.86 &
 -19.15 &
 -1.08 \\
\hline
 \( Z_{-} \)&
 -43.20 &
 -32.78 &
 -18.27 &
 -1.61  \\
\hline
\end{tabular}

\caption{\protect\( z\protect \) and \protect\( b\protect \) coefficients
for one loop matching factors for some two- and four-fermion operators,
for thin-link clover fermions with \protect\( C_{SW}=1\protect \),
and gauge action as labelled. Errors are \protect\( \pm 1\protect \)
in the last digit shown. }

\label{tab:currcl}
\end{table}

\begin{table}
\begin{tabular}{|c|c|c|c|c|}
\hline 
&
 Wilson &
Tree level Sym&
 Iwasaki &
 DWB2 \\
\hline
 \( Z_{V} \), \( Z_{A} \)&
 -14.80 &
 -11.24 &
 -6.76 &
 -2.63 \\
\hline 
 \( Z_{P} \), \( Z_{S} \)&
 -39.24 &
 -30.35 &
 -17.99 &
 -4.38 \\
\hline
 \( Z_{+} \)&
 -25.18 &
 -19.25 &
 -12.54 &
 -7.86 \\
\hline 
 \( Z_{-} \)&
 -68.06 &
 -51.45 &
 -29.0 &
 -5.35  \\
\hline
\end{tabular}

\caption{\protect\( z\protect \) and \protect\( b\protect \) coefficients
for one loop matching factors for some two- and four-fermion operators,
for naive fermions with \protect\( C_{SW}=1\protect \), and gauge
action as labelled. Errors are \protect\( \pm 1\protect \) in the
last digit shown.}

\label{tab:currnai}
\end{table}

\begin{table}
\begin{tabular}{|c|c|c|c|c|}
\hline 
&
 Wilson &
Tree level Sym&
 Iwasaki &
 DWB2 \\
\hline 
 \( Z_{V} \)&
 -1.38 &
 -1.18 &
 -0.89 &
 -0.50 \\
\hline 
 \( Z_{A} \)&
 -1.30 &
 -1.11 &
 -0.84 &
 -0.48 \\
\hline 
 \( Z_{P} \)&
 0.04 &
 0.54 &
 1.55 &
 3.77 \\
\hline 
 \( Z_{S} \)&
 -0.12 &
 0.41 &
 1.45 &
 3.72 \\
\hline 
 \( Z_{+} \)&
 -6.43 &
 -6.14 &
 -5.89 &
 -6.12 \\
\hline 
 \( Z_{-} \)&
 2.16 &
 3.11 &
 4.84 &
 8.34  \\
\hline
\end{tabular}

\caption{\protect\( z\protect \) and \protect\( b\protect \) coefficients
for one loop matching factors for some two- and four-fermion operators,
for clover fermions with \protect\( C_{SW}=1\protect \), HYP-smeared
links, and gauge action as labelled. Errors are \protect\( \pm 1\protect \)
in the last digit shown.}

\label{tab:currclh}
\end{table}

\begin{table}
\begin{tabular}{|c|c|c|c|c|}
\hline 
&
 Wilson &
Tree level Sym&
 Iwasaki &
 DWB2 \\
\hline 
 \( Z_{V} \), \( Z_{A} \)&
 -0.95 &
 -0.79 &
 -0.59 &
 -0.33 \\
\hline 
 \( Z_{P} \), \( Z_{S} \)&
 -0.62 &
 -0.01 &
 1.20 &
 3.61 \\
\hline 
 \( Z_{+} \)&
 -4.75 &
 -4.65 &
 -4.75 &
 -5.53 \\
\hline 
 \( Z_{-} \)&
 1.92 &
 2.95 &
 4.81 &
 8.37  \\
\hline
\end{tabular}

\caption{\protect\( z\protect \) and \protect\( b\protect \) coefficients
for one loop matching factors for some two- and four-fermion operators,
for naive fermions with \protect\( C_{SW}=1\protect \), HYP-smeared
links, and gauge action as labelled. Errors are \protect\( \pm 1\protect \)
in the last digit shown.}

\label{tab:currnaih}
\end{table}

\subsection{On-shell Scattering Amplitudes}

Let us recall that the motivation for introducing smeared links into
staggered fermions was to suppress the coupling between the region
of the fermion Brillouin zone near \( k_{\mu }=(0,0,0,0) \) and regions
of the Brillouin zone corresponding to doublers: one or more \( k_{\mu }\simeq \pi  \).
Smearing amounts to a form factor which suppresses the emission or
absorption of gluons whose exchange could scatter a quark from a \( k_{\mu }\simeq 0 \)
into a doubler state. The absence of this kind of scattering means
an improvement in flavor symmetry for staggered fermions since a process
which transforms a quark of one flavor (living in one part of the
Brillouin zone) into another flavor is reduced.

To quantify this scenario, let's imagine (in the continuum, first)
the scattering of two on-shell quarks of momentum \( \pm p_{1} \)
into two quarks of momentum \( \pm p_{2} \). The \( T- \)matrix
for the scattering is \begin{equation}
T(k)=[\bar{u}(p_{2})\gamma _{\mu }u(p_{1})]G_{\mu \nu }(k)[\bar{u}(-p_{2})\gamma _{\nu }u(-p_{1})]
\end{equation}
 where of course \( k=p_{1}-p_{2} \). Imposing the on-shell constraint
\( \gamma \cdot p|u(p_{i}\rangle =0 \), we see that the gauge-dependent
term (proportional to
$(\xi-1)k_\mu k_\nu/k^2$) vanishes,
 leaving the scattering amplitude proportional to the
Feynman gauge gluon propagator (and spinor factors). The reader can
quickly confirm that the same result obtains for naive fermions and
the Wilson gluon propagator.

\begin{figure}
{\centering \resizebox*{10cm}{!}{\includegraphics{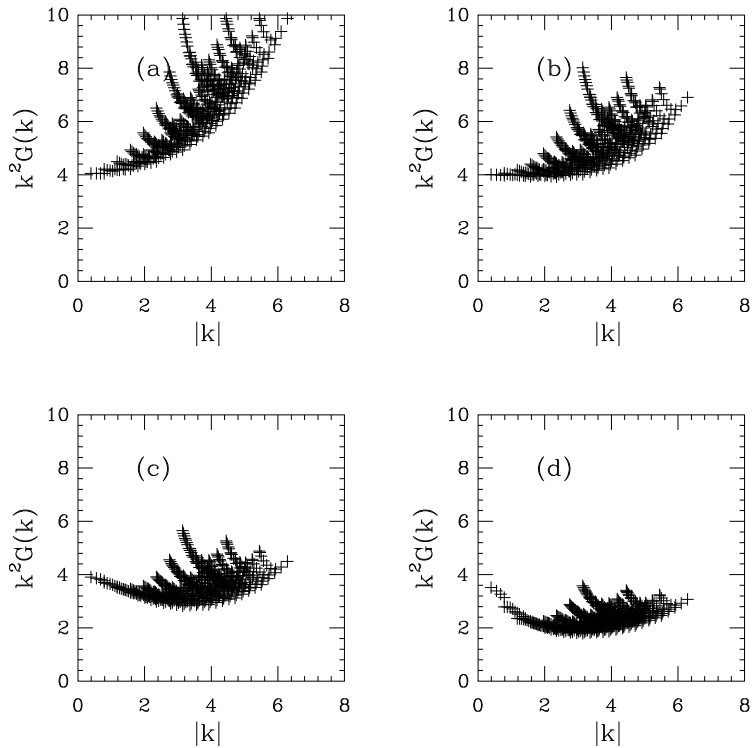}} \par}

\caption{\protect\( k^{2}G_{\mu \mu }(k)\protect \) for several lattice actions:
(a) Wilson (\protect\( c_{1}=0\protect \)), (b) tree level Symanzik
(\protect\( c_{1}=-1/12\protect \)), (c) Iwasaki (\protect\( c_{1}=-0.331\protect \))(d)
DWB2 (\protect\( c_{1}=-1.4088\protect \)). }

\label{fig:scatt}
\end{figure}

Now recall that the effect of smearing is to replace the gluon propagator
by the {}``smeared gluon propagator,{}'' Eq. \ref{eq:GEFF}. As
a way of comparing the level of flavor symmetry violation, we 
just look at the Feynman gauge propagator (appropriate for flavor
symmetry restoration with thin link fermions) or the smeared propagator.
We plot \( k^{2}\sum _{\mu }G_{\mu \mu } \) versus \( \sqrt{k^{2}} \)
for a set of momenta \( k_{\mu }=\pi n_{\mu }/8 \), \( n_{\mu }=1\dots 8 \).
Fig. \ref{fig:scatt} shows our results for thin link actions. As
\( c_{1} \) becomes more negative, the ``scattering amplitude{}''
at large \( k \) decreases. We would anticipate that flavor symmetry
would be improved by negative \( c_{1} \). We will confirm this expectation
in Sec. 6.

\begin{figure}
{\centering \resizebox*{10cm}{!}{\includegraphics{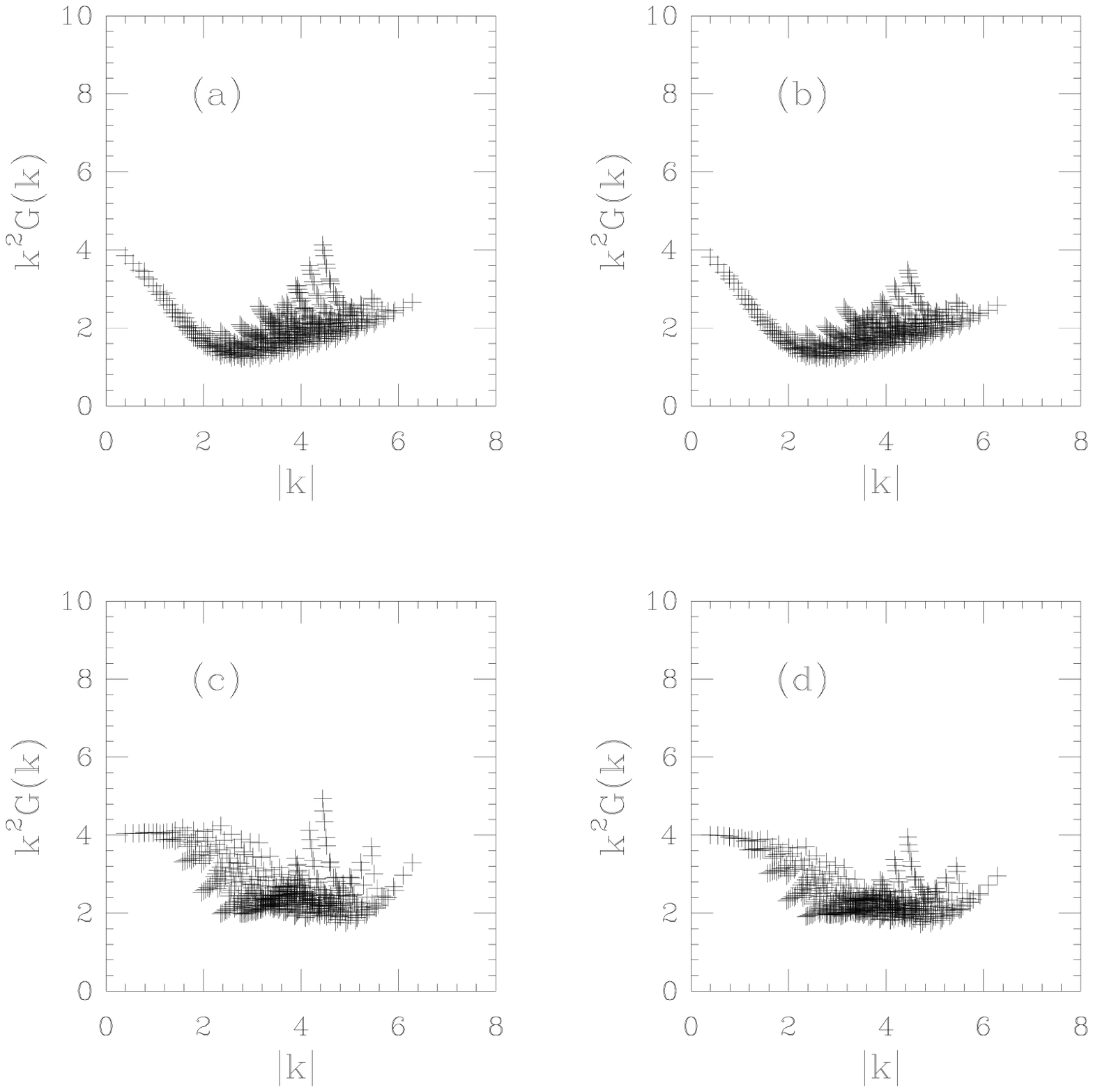}} \par}

\caption{\protect\( k^{2}{\cal G}_{\mu \mu }(k)\protect \) for two lattice actions
and two smearing functions: (a) Wilson (\protect\( c_{1}=0\protect \))
gauge action with HYP blocking, (b) tree level Symanzik (\protect\( c_{1}=-1/12\protect \))
gauge action with HYP blocking, (c) Wilson (\protect\( c_{1}=0\protect \))
gauge action with Asqtad blocking, (d) tree level Symanzik (\protect\( c_{1}=-1/12\protect \))
gauge action with Asqtad blocking, }

\label{fig:scatth}
\end{figure}

However, much greater suppression at large \( k \) is achieved by
converting to a smeared link. In Fig. \ref{fig:scatth} we show several
examples of \( k^{2}\sum _{\mu }{\mathcal{G}}_{\mu \mu } \), which
parameterizes the scattering of smeared link fermions. HYP links and
the Wilson gauge action produce more suppression than thin links and
the DWB2 action. Clearly, a combination of greater \( c_{1} \) and
a smeared link would produce a larger effect.

There is an obvious qualitative connection between Figs. \ref{fig:scatt}
and \ref{fig:scatth} and the results of one-loop perturbation theory
for matching coefficients:
As the magnitude of the gluon propagator shrinks at large \( k \),
so does its contribution to the integrals of the perturbative calculation.

\section{The non-perturbative static potential}

We start our discussion of the non-perturbative properties of the
different actions with the static quark potential. The perturbative
results of Sect. 3.2 suggest that the thin link Symanzik action has
the smallest lattice distortion. The Wilson action has a positive
lattice correction at small distance and observable rotational symmetry
breaking even at \( r/a=2-3 \). The perturbative DBW2 potential has
a larger and negative correction at \( r/a=1 \) and observable rotational
symmetry breaking even at \( r/a=4-5 \). The perturbative HYP smeared
Wilson gauge action potential has similar distortion at \( r/a=1 \)
as the DBW2 but much smaller rotational symmetry violation while the
Asqtad smeared Wilson gauge potential has smaller distortion at short
distances but larger rotational symmetry violation. 
\begin{figure}
{\centering \resizebox*{15cm}{!}{\includegraphics{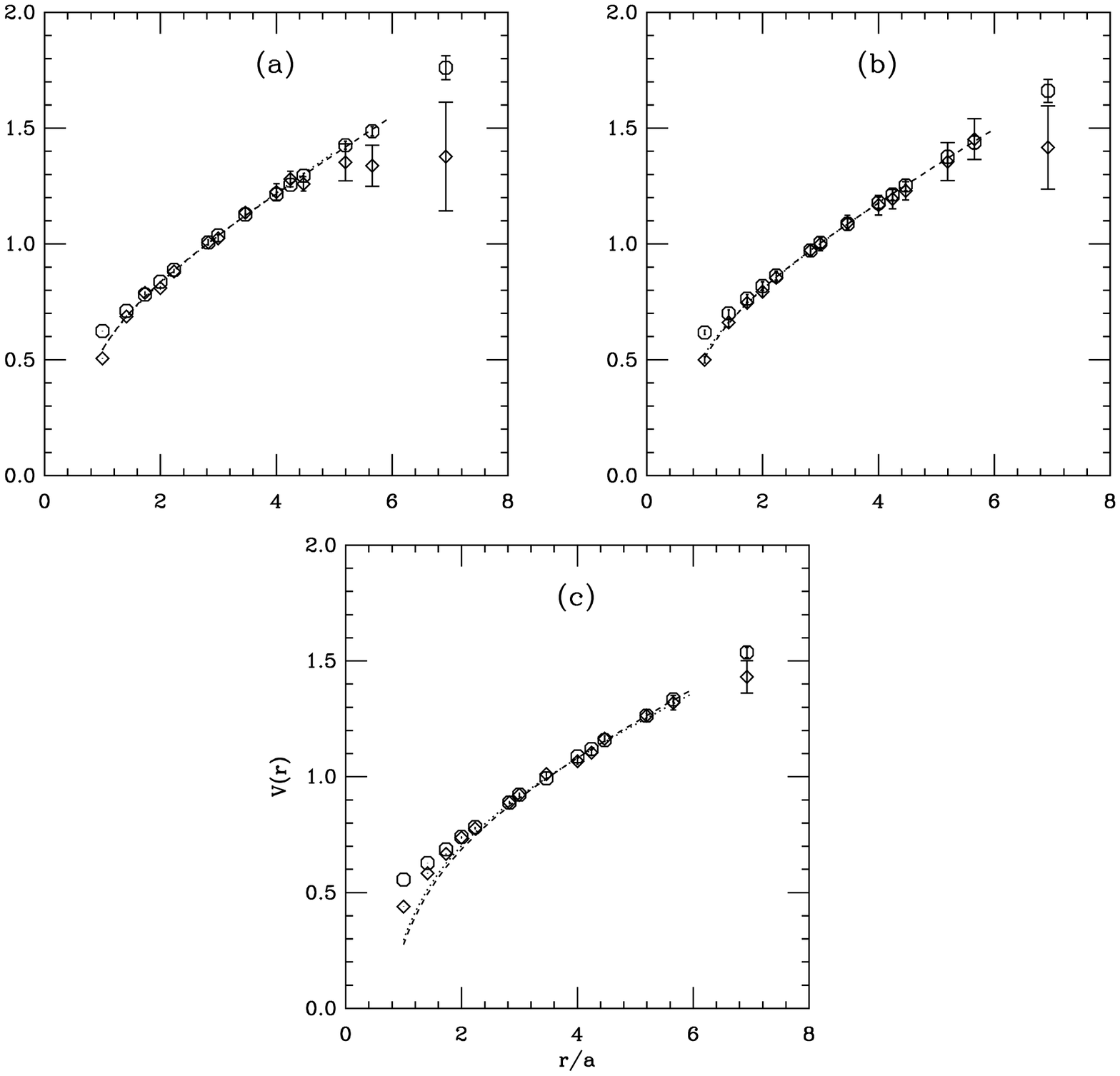}} \par}

\caption{The static potential measured with a) Wilson gauge action, b) 1-loop
Symanzik gauge action and c) DBW2 gauge action. In all cases
both the thin link (diamonds) and HYP smeared (octagons) potentials
are plotted, shifted to agree at \protect\( r/a=\sqrt{7}\protect \).
The dotted and dashed lines are the fitted continuum potentials as
described in the text. \label{fitpot}}
\end{figure}

We studied the non-perturbative static potential on Wilson, 1-loop
Symanzik and DBW2 gauge backgrounds with and without HYP smearing.
The simulations were carried out on \( 8^{3}\times 24 \) lattices
with Sommer scale \( r_{0}/a\sim 3.0 \) for all three actions, at \( \beta =5.7 \)
for Wilson, \( \beta =7.775 \) for Symanzik and \( \beta =0.82426 \)
for the DBW2 action. At large distances the smeared and thin link
potentials differ only by an irrelevant additive constant. In Fig.
\ref{fitpot} we plot both the smeared and thin link potentials, the
former one shifted by a constant to match the thin link potential
at \( r/a=\sqrt{7} \), a somewhat arbitrarily chosen matching point.
We fit each potential following the method proposed in \cite{Potential:Edwards:1998xf}
and used with the HYP potential in \cite{Hasenfratz:2001tw, Gattringer:2001jf}
taking a four parameter functional form \[
\label{V_latt}
V_{\rm {latt}}(r)=V_{\rm {cont}}(r)+\epsilon (V_{c}(r)-\frac{1}{4\pi r})\]
 where \( V_{\rm {cont}} \) is the continuum potential\[
\label{V_cont}
V_{\rm {cont}}=-\frac{e}{r}+V_{0}+\sigma r,\]
 and \( V_{c}(r) \) is the lattice Coulomb potential of Eq. \ref{eq:pert_pot}.
The term \( \epsilon (V_{c}(r)-1/(4\pi r) \)) is an attempt
to model and remove the lattice artifacts of the potential. It is
difficult to judge how much of the lattice artifacts can be described
by this term, only the quality of the fit can justify its use. 

Fig. \ref{fitpot}a shows the Wilson gauge action potential measured
with thin links (diamonds) and with HYP links (octagons). The dotted
line of the figure correspond to the fit of the continuum potential
\( V_{\rm {cont}} \) after the removal of the lattice artifacts while
the dashed line is the same continuum potential obtained with the
HYP smeared links and their corresponding perturbative corrections.
The fact that it is impossible to resolve the two different lines
in the figure indicates that the lattice artifacts are consistently removed.
Also, the sign and relative magnitude of the lattice corrections are
what we expected from the perturbative Coulomb potential. Fig. \ref{fitpot}b
is the same for the 1-loop Symanzik action. The smeared and thin
link results are consistent, indistinguishable. The thin link potential has very small
lattice artifacts but after removing the lattice correction both thin
and HYP potentials predict the same continuum values. In Fig. \ref{fitpot}c
we plot the corresponding potential data for the DBW2 action. The
agreement between the thin and HYP smeared potentials is good though not as perfect
as in the previous two cases, mainly because of the stronger rotational
symmetry violation of the thin link potential.The results
agree with the perturbative predictions: even the thin link DBW2 potential
has a large distortion at \( r/a=1 \), about the same as the HYP
link potential with Wilson or 1-loop Symanzik action. If the value
of the potential (or any other quantity) is important at \( r/a=1 \),
the DBW2 action is not a good choice to use. 

It is difficult, if not impossible, to prove how the lattice Coulomb term 
could describe the lattice artifacts of the non-perturbative potential. 
Only the fact that  the different potential measurements with thin and 
HYP links give consistent continuum results for all three gauge 
actions justifies its use.

Recent calculations of the static potential with the DBW2 action did
not show such a large distortion at small distances. However in 
Ref. \cite{Necco:2002zy}
only the on-axis potential was measured at distances \( r/a\geq 2 \).
From that data it would have been hard to see the distortion at small
distances.

\section{Topology and the gauge action}

\subsection{Smooth instantons}

It is generally believed that flavor symmetry violation for staggered
fermions, the residual chiral symmetry violation of domain wall fermions,
and the computationally most demanding small eigenmodes of the overlap
fermions are due to small scale, plaquette level vacuum fluctuations.
Smearing attempts to remove these objects by averaging the gauge links
while the Iwasaki and DBW2 actions do the same by making it energetically
unfavorable to create them in the first place. That mechanism can
be seen clearly from the action of smooth instantons. 
\begin{figure}
{\centering \resizebox*{15cm}{!}{\includegraphics{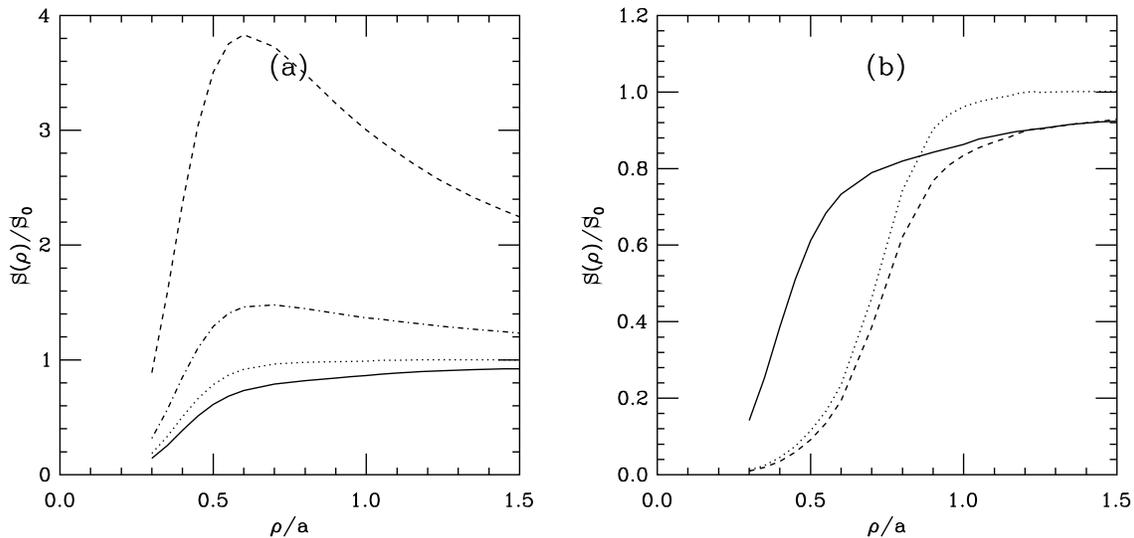}} \par}

\caption{The action of smooth instantons, normalized by the continuum value, 
as the function of the instanton radius, calculated with different
actions. a) solid line: Wilson action, dotted line: tree level Symanzik
action; dashed-dotted line: Iwasaki action; dashed line: DBW2 action.
b) solid line: Wilson action (thin link); dashed line: HYP smeared
Wilson ation; dotted line: HYP smeared tree level Symanzik action.\label{fig:instprof}}
\end{figure}
We calculated the instanton action on a set of smooth instantons with
varying radii. These instantons were created in singular gauge 
on \(32^4\) lattices and blocked twice 
in order to approximate the smooth, continuum solution.
 In Fig. \ref{fig:instprof}a we show the result,
normalized by the continuum instanton action, for Wilson, tree level
Symanzik, Iwasaki and DBW2 actions. A perfect lattice action should
have a profile that is less than one for small radii and one for radii
larger than a critical value where the vacuum fluctuation becomes
an instanton. At the critical radius the action function is non-analytic,
it develops a kink. None of our actions can reproduce this behavior,
but not surprisingly the closest to it is the tree level Symanzik
action. The Wilson action profile approaches the continuum value more slowly,
from below. The Iwasaki action overshoots the continuum value by about
50\% at \( r/a\sim 0.6 \), suppressing fluctuations of this size.
The correponding DBW2 curve is nothing less than shocking. The curve
rises to almost four at \( r/a\sim 0.6 \) and even at distance \( r/a=1.5 \)
it is above two. The DBW2 action strongly suppresses instantons and
dislocations with radius \( 0.3<r/a<2-3 \). The very small fluctuations
are still present, but small radius instantons are disfavored.  On
lattices where these small instantons are important physically one
would expect fairly large lattice artifacts from the DBW2 action. 

Smearing attempts to remove dislocations seen by the fermions by averaging
the gauge links. Fig. \ref{fig:instprof}b compares the HYP smeared
instanton profiles of the Wilson and Symanzik actions. For reference
we include the thin link Wilson action profile again. The most important
conclusion from Fig. \ref{fig:instprof}b is that smearing removes
dislocations with radius \( r/a<0.5 \), the action profile rises only
at \( r/a\sim 0.7 \). Even though the gauge configuration can
have plenty of small instantons and dislocations, most of these are
not seen by the fermions.

\subsection{The autocorrelation of the topological charge }

\vspace{3mm}
\begin{figure}
{\centering \resizebox*{10cm}{!}{\includegraphics{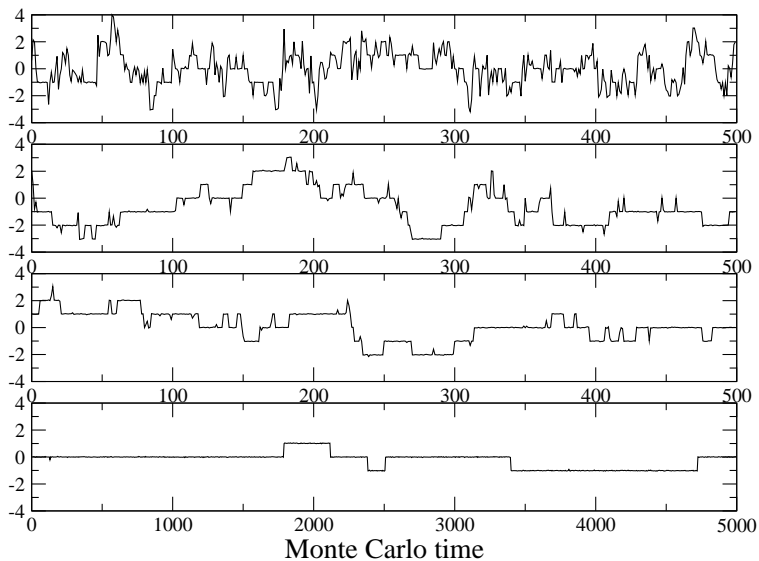}} \par}

\caption{The Monte Carlo time history of the topological charge with various
gauge actions. The actions are (from top to bottom)
Wilson (\protect\( \beta =6.0\protect \)),
Symanzik (\protect\( \beta =8.4\protect \)),
Iwasaki (\protect\( \beta =2.6\protect \))
and DBW2 (\protect\( \beta =1.04\protect \)).
\label{fig:Q_hist}}
\end{figure}
We have seen that as the coefficient \( c_{1} \) in the action becomes
more negative, the action favors small instantons less and less. In
a Monte Carlo simulation with local (one-link) updates, change of
topology always occurs through the (dis)appearance of small topological
objects. It is thus not very surprising that the suppression of small
topological objects implies that topology changes less often. In Fig. \ref{fig:Q_hist}
we show the Monte Carlo time history of the topological charge with
different gauge actions. The charge was measured using the RG improved
charge operator \cite{DeGrand:1997gu, Hasenfratz:1998qk} after 8 levels of APE
smearing steps. The units on the horizontal axis correspond
to ten full sweeps of a combination of one overrelaxation and one
Metropolis step over the whole lattice. The lattice size in these
simulations was \( 12^{4} \) in all cases and the \( \beta  \) values
were matched to correspond to the same lattice spacing, \( a=0.095 \)
fm, set by the Sommer scale. Indeed, the difference among the gauge
actions is striking. Notice that for the DBW2 action the MC time scale
is an order of magnitude different. The integrated autocorrelation
time of the topological charge was estimated to be 100 and 700 sweeps
for the Wilson and the Iwasaki action respectively. In the case of
the DBW2 action the autocorrelation time is so enormous that the available
data was not  enough even to estimate it. It is also interesting to
observe that the plaquette autocorrelation time is 10, 7 and 5.5 for
the Wilson, Iwasaki and the DBW2 action. The very large autocorrelation
time of the topological charge of the DBW2 action was also noted by
the RBC collaboration \cite{Orginos:1998ue}.

It is also noticeable that as \( c_{1} \) becomes more negative,
and the change between topological sectors occurs less frequently,
the charge also becomes closer  on the average to integer values. This
also indicates the strong suppression of small instantons and in general
the suppression of gauge configurations close to the boundary between
two charge sectors.

\subsection{The instanton size distribution}

The instanton action profiles of Fig. \ref{fig:instprof} indicate
a slight suppression of small instantons for the Iwasaki action, and
strong suppression of small and even larger instantons for the DBW2
action. On configurations with \( a\sim0 .095 \) fm the average instanton
radius is about \( r/a=3 \) but smaller instantons should be also
present. In order to see if the different gauge actions have different
instanton size distributions we have measured instanton sizes on a
set of \( a\sim0 .095 \) fm configurations. We measured the topological
charge density after 2-4 levels of HYP smearing \cite{Hasenfratz:2001wd} and
compared it to smooth instanton profiles. This is the same method
we used in Ref. \cite{DeGrand:1997gu, Hasenfratz:1998qk} except we
do not extrapolate the instanton size to zero smearing level.
\begin{figure}
{\centering \resizebox*{15cm}{!}{\includegraphics{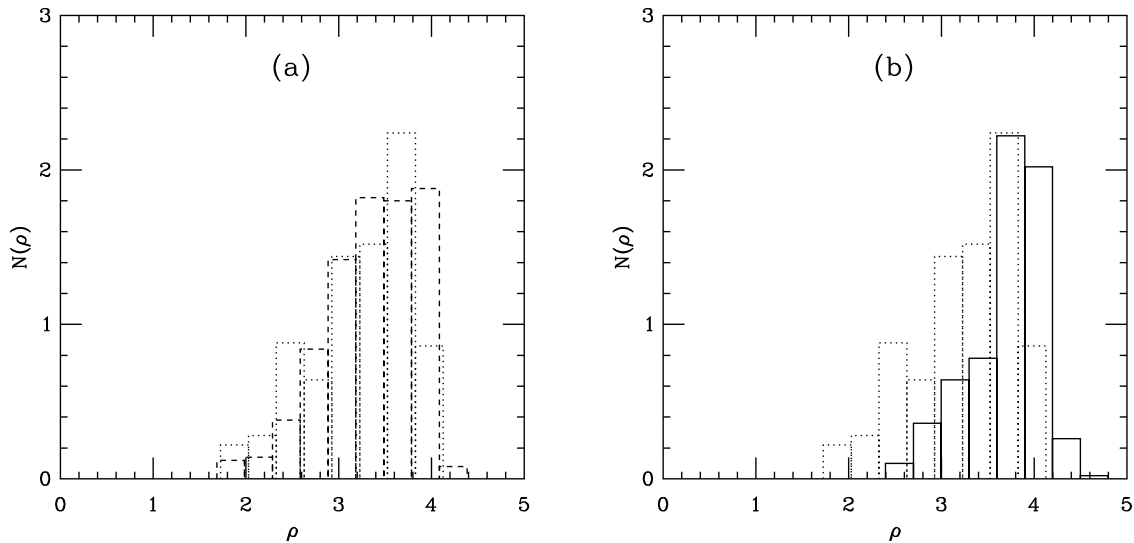}} \par}

\caption{Instanton size distribution of the different actions after two HYP
smearing steps. a) Wilson (dotted lines) and Iwasaki (dashed lines)
actions; b) Wilson (dotted lines) and DBW2 (solid lines) actions.
\label{fig:instsize}}
\end{figure}
In Fig. \ref{fig:instsize} we compare the instanton size distribution
of the different actions after two HYP smearing steps. We have used
the same \( 12^{4} \), \( a=0.095 \) fm configurations we analyzed
in the previous section and normalized the distribution by the number
of configurations. As Fig. \ref{fig:instsize}a illustrates, there
is not much difference between the Wilson and Iwasaki actions. The
size distribution peaks around \( \rho /a=3.5 \) and both actions
predict the same topological density. (We are not concerned about
the physical significance of the topological density here.
 Since we have used the
same lattice spacing and analysis method with both configuration sets,
comparing the two densities gives information about the two actions.)
In contrast, the Wilson and DBW2 actions differ significantly, as
Fig. \ref{fig:instsize}b illustrates. The smaller instantons are
suppressed by the DBW2 action, and the topological density is about 30\% less
on the DBW2 than on the Wilson configurations. If we consider
the physical picture of quark propagation, where the quarks hop from
instanton to anti-instanton in the vacuum \cite{Schafer:1998wv}, lack
of instantons could point to observable scaling violations in the
light hadron spectrum.

\section{Flavor symmetry violation in staggered actions}

Smeared actions are used with staggered fermions because they considerably
reduce flavor symmetry violations. Both the Asqtad and HYP smearing
are \( O(a^{2}) \) perturbative improved though the coefficients
of the HYP smearing are non-perturbatively optimized. Relative to
the thin link staggered action, Asqtad fermions improve flavor symmetry
by a factor of five, HYP fermions by about a factor of ten. Based
on our perturbative and instanton analyzes, flavor symmetry could
also be improved by modifying the gauge action. At first this approach
might look attractive: it is much easier to simulate thin link fermions
with a complicated gauge action than smeared link fermions. However
taking the easy way might have serious consequences later on. The
potential data indicates that the DBW2 action distorts short distance
behavior, the time evolution of the topological charge points to unacceptably
long autocorrelation times, the low topological density could imply
large lattice artifacts. Nevertheless, in this section we consider
the possibility of using different gauge actions with and without
smearing in staggered fermion simulations and investigate the level
of flavor symmetry violation these actions show.

 We have studied the
quenched staggered spectrum on our \( 8^{3}\times 24 \), \( r_{0}/a\sim 3.0, \)
\( a\sim 0.17 \) fm configurations. Before presenting our results
for the spectrum, first we look at the distribution of the plaquette
on these configurations. In \cite{Hasenfratz:2001hp} it was argued
and illustrated that the end tail of the plaquette distribution is
correlated with flavor symmetry breaking of staggered fermions. The
argument is quite simple: flavor symmetry violation is caused by the
strongly fluctuating gauge fields at the hypercubic level. These gauge
links create plaquettes with very small value, therefore the number of plaquettes
with very small value indicate the level of flavor symmetry breaking
the fermions observe. 
\begin{figure}
{\centering \resizebox*{15cm}{!}{\includegraphics{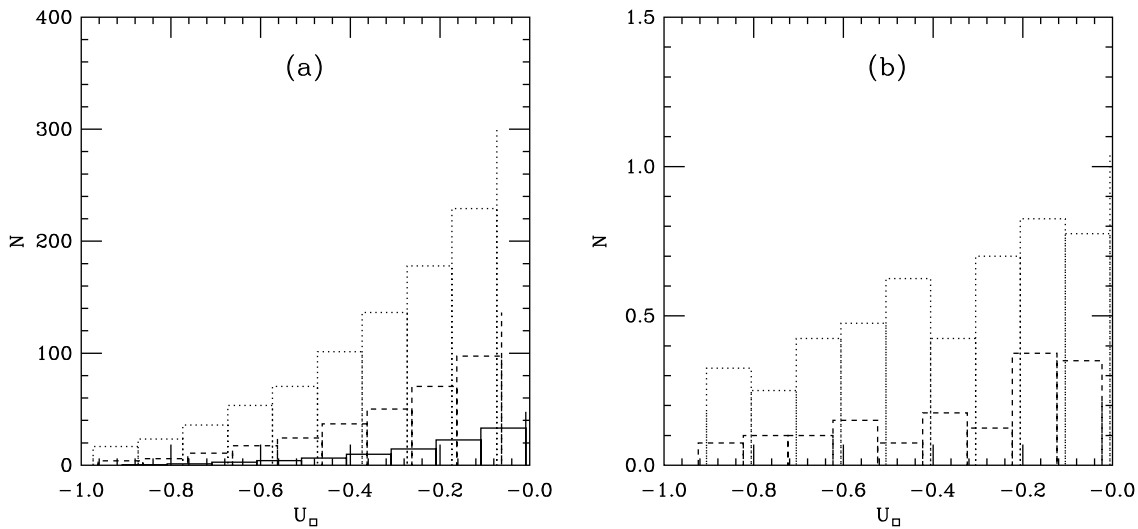}} \par}

\caption{The tail of the plaquette distribution for different gauge actions
a) with thin links and b) after HYP smearing. Dotted line: Wilson,
dashed line: 1-loop Symanzik, solid line: DBW2 action. Observe the
scale difference of the two figures.\label{fig:plaq_distr} }
\end{figure}

In Fig. \ref{fig:plaq_distr} we show the tail of the plaquette
distribution, normalized by the number of configurations for the Wilson,
1-loop Symanzik and DBW2 gauge actions. Fig. \ref{fig:plaq_distr}a
compares the plaquettes constructed from thin links. Not surprisingly,
the DBW2 action is a factor of eight better than the Wilson action.
The 1-loop Symanzik action is also better than the Wilson action by
about a factor of two. In Fig. \ref{fig:plaq_distr}b we plot
the plaquette distribution of the Wilson and 1-loop Symanzik actions
after one level of HYP blocking. The latter is again a factor of two
better than the former, but both are an order of magnitude better
than the thin link DBW2 action. (Observe the scale difference of the
two figures.)
\begin{figure}
{\centering \resizebox*{10cm}{!}{\includegraphics{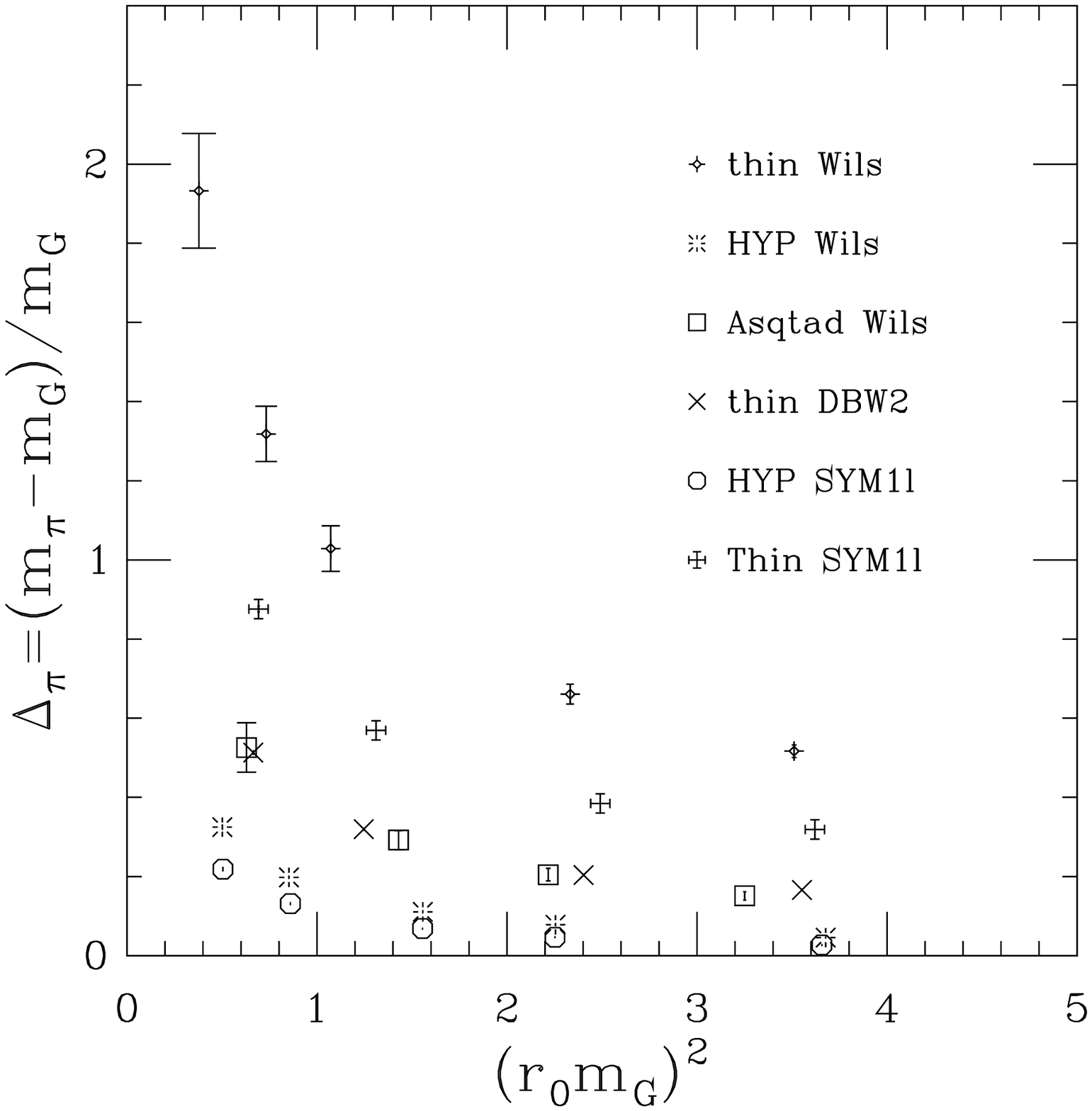}} \par}

\caption{Flavor symmetry violation of staggered fermions on different 
gauge action backgrounds with different smearing transformations. 
All data on this plot is from quenched simulations with lattice spacing \(a \simeq 0.17 \)fm.  \label{fig:dpi_vs_mG}}
\end{figure}
 Does the implication from the tail of the plaquette agree with the
actual flavor symmetry violation of the different actions? We use
the parameter \begin{equation}
\label{eq:delta-pi}
\Delta _{\pi }=\frac{m_{\pi }-m_{G}}{m_{G}},
\end{equation}
which measures the relative difference between the Goldstone pion
mass and the non-Goldstone pions, to compare the different actions.
This quantity diverges at zero quark masses and depends strongly on
the lattice spacing, but since we have done all the simulations at
approximately identical lattice spacings and volumes, \( \Delta _{\pi } \)
as a function of the Goldstone particle, is a good indicator of
the flavor symmetry violation of the different actions. The results,
shown in Fig. \ref{fig:dpi_vs_mG}, confirms these expectations. The
thin link 1-loop Symanzik action improves flavor symmetry relative
to the Wilson action by about 30\%. The thin DBW2 action is even better,
it has flavor symmetry violation at the level of the Asqtad smeared
Wilson action. HYP smearing, even on Wilson configurations, is a
factor of two better. One can reduce flavor symmetry violation even
further by using the 1-loop Symanzik action. It is not easy to see
from the figure, but 1-loop Symanzik is about 30\% better than Wilson,
even after HYP smearing. Since the Symanzik action has better scaling
scaling properties than the Wilson action and does not suffer from
topological autocorrelation
slow down like the DBW2 and Iwasaki actions do, a HYP
smeared Symanzik action appears to be the best choice for staggered
simulations.

\section{Smearing, improved actions and the overlap operator}

\subsection{Fermionic charge and overlap}
Since a large negative $c_1$ suppresses the creation of small instantons
or other objects where the gauge field is on the boundary between
different topological sectors, gauge actions with a more
negative \( c_{1} \) result in smaller residual masses in domain
wall simulations. A similar mechanism is at work in the case of the
overlap. The overlap operator is defined in terms of a simple Dirac
operator (e.g.\ Wilson) \( D_{0} \) by the formula 
\begin{equation}
D_{ov}=1-A\left[ A^{\dagger }A\right] ^{-\frac{1}{2}},\hspace {1cm}A=1+s-D_{0},
\end{equation}
 where \( s \) is a real parameter. If the gauge configuration is
close to the boundary between different
 topological sectors, \( A^{\dagger }A \)
has to have a small eigenvalue. Therefore, the suppression of these
boundary gauge configurations can also thin out the small eigenvalues
of \( A^{\dagger }A \). This is  important for practical
applications since the cost
of the overlap is governed by the condition number of \( A^{\dagger }A \).
Assuming for instance that Chebyshev polynomials are used to approximate
the inverse square root, the order of the Chebyshev polynomial is
inversely proportional to the square root of the smallest eigenvalue
where the Chebyshev approximation has to work. In Table~\ref{tab:smallev}
we show the average smallest and eighth smallest eigenvalue
of \( A^{\dagger }A \) with \( D_{0} \) being the Wilson Dirac operator, on 
different sets of gauge backgrounds with the same lattice spacing. The lattice
size was $12^4$ and the $\beta$ values were 6.0 (Wilson), 2.60 (Iwasaki),
8.40 (Symanzik) and 1.04 (DBW2) and each set contained 50 independent 
configurations. For this qualitative 
test we fixed the value of $s$ to be 0.5, except on the HYP smeared 
configurations, where it was set to zero. Although in principle $s$ would
have to be optimized for each type of gauge background separately,
we chose to fix it close to the overall optimal value. This is 
sufficient for our purposes, moreover, an optimization for the
smallest eigenvalues of $A^\dagger A$ and for locality would yield different 
values. 

There is a clear trend that a more negative value of 
\( c_{1} \) pushes up the smallest eigenvalues.
However, the reader's attention is called to the last entry in the
table, where we have HYP-smeared the links in the fermion action
and retained the Symanzik gauge action.  The gain in time
for simulating this action is almost a  factor of two better than for the DBW2
action.

Generally, overlap simulations are accelerated by projecting out
(and treating exactly) the eigenvectors corresponding to the smallest
few eigenvectors. This results in a gain of about a factor of 2 for the Wilson
action, and smaller factors for the other actions studied. To facilitate
a comparison we also included in the table the factor one can gain 
in speed compared to Wilson gauge action without projecting out any
eigenvector.

\begin{table}
{\centering \vspace{3mm}\par}

{\centering \begin{tabular}{|l|r|r|r|r|}
\hline 
Action &
 \( \langle \lambda _{1}\rangle  \)&
 \( t_{W1}/t \)&
 \( \langle \lambda _{8}\rangle  \)&
 \( t_{W1}/t \)\\
\hline
Wilson &
 0.013(2) &
 1.0 &
 0.061(1) &
 2.17 \\
\hline
Symanzik &
 0.044(4) &
 1.84 &
 0.105(1) &
 2.84 \\
\hline
Iwasaki &
 0.065(4) &
 2.24 &
 0.130(1) &
 3.16 \\
\hline
DBW2 &
 0.160(5) &
 3.51 &
 0.217(1) &
 4.09 \\
\hline
Symanzik+HYP &
 0.46(3) &
 5.95 &
 0.737(2) &
 7.52  \\
\hline
\end{tabular}\par}

\caption{The average smallest eigenvalue \protect\( (\lambda _{1})\protect \)
and the 8th smalles eigenvalue \protect\( (\lambda _{8})\protect \)
of \protect\( A^{\dagger }A\protect \) and the factor of CPU time
reduction compared to the Wilson action with no eigenvector projected out. 
The first two columns refer
to the case when no eigenvalues are projected out, the last to columns
to the one when the eight lowest eigenvectors are projected out and
treated exactly. \label{tab:smallev}}
\end{table}

A major difference between the overlap and domain wall formulation
is that for the overlap small eigenvalues of \( A^{\dagger }A \)
are only a nuisance, as they make the calculation more expensive.
On the other hand, in domain wall simulations the extension of the
lattice in the fifth direction is fixed and this results in different
chirality violations and residual mass effects configuration by configuration.
This is the main reason why improvement of the gauge action 
was so badly needed in the case of DW fermions. As our results with
the overlap suggest, improving the Domain Wall technology might
be a better solution than going to extremes in tuning the gauge action.

\subsection{Localization}

\begin{figure}
{\centering \resizebox*{10cm}{!}{\includegraphics{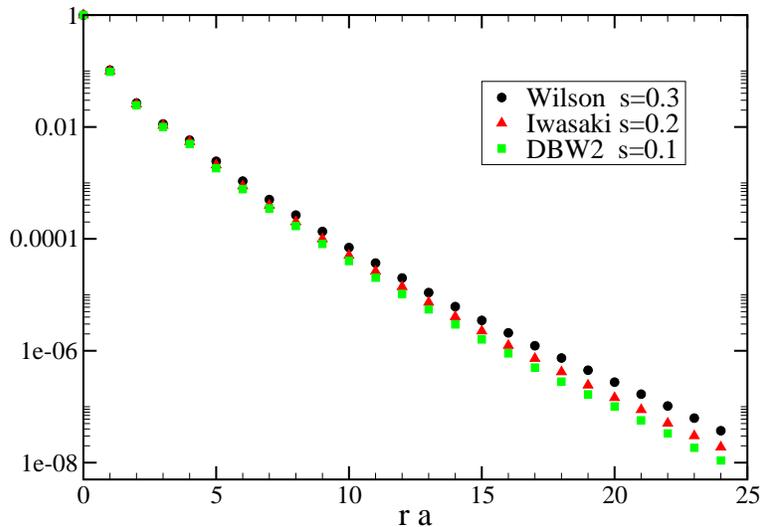}} \par}
\caption{The localization of the overlap in gauge backgrounds generated with
various gauge actions. The parameter \protect\( s\protect \) was
chosen to make the operator as local as possible in the given gauge
background.\label{loc_vargauge}}
\end{figure}

We saw that the density of low modes of $A^\dagger A$ depends on the
gauge action. There have been speculations on the connection between
low modes of $A^\dagger A$ and the locality of the overlap 
\cite{Hernandez:1998et,Kovacs:2002nz}. It is therefore interesting
to compare the locality of the overlap in the different gauge 
backgrounds studied in the present work. This is done in Fig.\ 
\ref{loc_vargauge}, where the quantity
\begin{equation}
  f(r) = \max \left\{ ||D_{ov}\psi (x)|| : \sum_\mu|x_\mu|=r \right\}
\end{equation}
is plotted as a function of the so called ``taxi driver distance'',
$r = \sum_\mu |x_\mu|$ from a localized source 
$\psi_k(x) = \delta(x) \delta_{kj}$. This quantity was introduced
in \cite{Hernandez:1998et} to measure the (non)-locality of the 
overlap operator.

There are no surprises here. As seen in the figure, the general trend is 
that a large reduction in the number of small modes of $A^\dagger A$ results 
in a slight improvement of the locality of the overlap, in accordance 
with the results of \cite{Kovacs:2002nz}, where different types and 
degrees of smearing were shown to have a similar effect.

\section{Conclusions}
In this paper we have investigated the reduction of lattice artifacts
which can be achieved by altering the gauge field self-interaction and the
fermion-gauge coupling.  We focussed mostly -- but not exclusively --
on the reduction of lattice chiral symmetry breaking artifacts.
Chiral improvement can be achieved by either altering the gauge action,
or the fermion-gauge field coupling, or by a combination.  That the
two alterations produce similar effects is most starkly revealed
by perturbation theory, where one sees that either effect alters
the effective gluon propagator in fermionic Feynman diagrams.
A variety of perturbative and nonperturbative tests reveal that
a large $c_1$ in the gauge action can improve flavor symmetry violation
for staggered fermions or the efficiency of implementing domain wall
fermions or the overlap.  However, a much more dramatic improvement can
be achieved by replacing the thin link variable in the fermion action with
smeared links.  And the use of a large $c_1$ in the gauge action
introduces a number of bad features into simulations: most notably
a distortion of the heavy quark potential at short distance and long
simulation time autocorrelations of the topological charge.

\section*{Acknowledgments}
Thw work of T.~D. and A.~H. was supported by the
U.~S. Department of Energy with grant
DE-FG03-95ER40894.
The work of T.~K. was
supported by the EU's Human Potential Program under
contract HPRN-CT-2000-00145, by Hungarian science grant OTKA-T032501,
and also partly by a Bolyai Fellowship.

\bibliographystyle{apsrev}
\bibliography{lattice}

\begin{thebibliography}{35}
\expandafter\ifx\csname natexlab\endcsname\relax\def\natexlab#1{#1}\fi
\expandafter\ifx\csname bibnamefont\endcsname\relax
  \def\bibnamefont#1{#1}\fi
\expandafter\ifx\csname bibfnamefont\endcsname\relax
  \def\bibfnamefont#1{#1}\fi
\expandafter\ifx\csname citenamefont\endcsname\relax
  \def\citenamefont#1{#1}\fi
\expandafter\ifx\csname url\endcsname\relax
  \def\url#1{\texttt{#1}}\fi
\expandafter\ifx\csname urlprefix\endcsname\relax\def\urlprefix{URL }\fi
\providecommand{\bibinfo}[2]{#2}
\providecommand{\eprint}[2][]{\url{#2}}

\bibitem[{\citenamefont{Toussaint}(2002)}]{Toussaint:2001zc}
\bibinfo{author}{\bibfnamefont{D.}~\bibnamefont{Toussaint}},
  \bibinfo{journal}{Nucl. Phys. Proc. Suppl.} \textbf{\bibinfo{volume}{106}},
  \bibinfo{pages}{111} (\bibinfo{year}{2002}),
  \eprint[http://arXiv.org/abs]{hep-lat/0110010}.

\bibitem[{\citenamefont{Hasenfratz and Knechtli}(2002)}]{Hasenfratz:2002jn}
\bibinfo{author}{\bibfnamefont{A.}~\bibnamefont{Hasenfratz}} \bibnamefont{and}
  \bibinfo{author}{\bibfnamefont{F.}~\bibnamefont{Knechtli}},
  \bibinfo{journal}{Comput. Phys. Commun.} \textbf{\bibinfo{volume}{148}},
  \bibinfo{pages}{81} (\bibinfo{year}{2002}),
  \eprint[http://arXiv.org/abs]{hep-lat/0203010}.

\bibitem[{\citenamefont{Lepage}(1999)}]{Lepage:1998vj}
\bibinfo{author}{\bibfnamefont{G.~P.} \bibnamefont{Lepage}},
  \bibinfo{journal}{Phys. Rev.} \textbf{\bibinfo{volume}{D59}},
  \bibinfo{pages}{074502} (\bibinfo{year}{1999}), \eprint{hep-lat/9809157}.

\bibitem[{\citenamefont{Orginos and Toussaint}(1999)}]{Orginos:1998ue}
\bibinfo{author}{\bibfnamefont{K.}~\bibnamefont{Orginos}} \bibnamefont{and}
  \bibinfo{author}{\bibfnamefont{D.}~\bibnamefont{Toussaint}}
  (\bibinfo{collaboration}{MILC}), \bibinfo{journal}{Phys. Rev.}
  \textbf{\bibinfo{volume}{D59}}, \bibinfo{pages}{014501}
  (\bibinfo{year}{1999}), \eprint{hep-lat/9805009}.

\bibitem[{\citenamefont{Orginos et~al.}(1999)\citenamefont{Orginos, Toussaint,
  and Sugar}}]{Orginos:1999cr}
\bibinfo{author}{\bibfnamefont{K.}~\bibnamefont{Orginos}},
  \bibinfo{author}{\bibfnamefont{D.}~\bibnamefont{Toussaint}},
  \bibnamefont{and} \bibinfo{author}{\bibfnamefont{R.~L.} \bibnamefont{Sugar}}
  (\bibinfo{collaboration}{MILC}), \bibinfo{journal}{Phys. Rev.}
  \textbf{\bibinfo{volume}{D60}}, \bibinfo{pages}{054503}
  (\bibinfo{year}{1999}), \eprint{hep-lat/9903032}.

\bibitem[{\citenamefont{Hasenfratz and Knechtli}(2001)}]{Hasenfratz:2001hp}
\bibinfo{author}{\bibfnamefont{A.}~\bibnamefont{Hasenfratz}} \bibnamefont{and}
  \bibinfo{author}{\bibfnamefont{F.}~\bibnamefont{Knechtli}},
  \bibinfo{journal}{Phys. Rev.} \textbf{\bibinfo{volume}{D64}},
  \bibinfo{pages}{034504} (\bibinfo{year}{2001}), \eprint{hep-lat/0103029}.

\bibitem[{\citenamefont{Iwasaki}(1983)}]{Iwasaki:1983ck}
\bibinfo{author}{\bibfnamefont{Y.}~\bibnamefont{Iwasaki}}
  (\bibinfo{year}{1983}), \bibinfo{note}{uTHEP-118}.

\bibitem[{\citenamefont{Ali~Khan et~al.}(2000)}]{Iwa:AliKhan:1999ib}
\bibinfo{author}{\bibfnamefont{A.}~\bibnamefont{Ali~Khan}} \bibnamefont{et~al.}
  (\bibinfo{collaboration}{CP-PACS}), \bibinfo{journal}{Nucl. Phys. Proc.
  Suppl.} \textbf{\bibinfo{volume}{83}}, \bibinfo{pages}{360}
  (\bibinfo{year}{2000}), \eprint[http://arXiv.org/abs]{hep-lat/9909075}.

\bibitem[{\citenamefont{Ali~Khan et~al.}(2001)}]{DW:AliKhan:2000iv}
\bibinfo{author}{\bibfnamefont{A.}~\bibnamefont{Ali~Khan}} \bibnamefont{et~al.}
  (\bibinfo{collaboration}{CP-PACS}), \bibinfo{journal}{Phys. Rev.}
  \textbf{\bibinfo{volume}{D63}}, \bibinfo{pages}{114504}
  (\bibinfo{year}{2001}), \eprint[http://arXiv.org/abs]{hep-lat/0007014}.

\bibitem[{\citenamefont{Takaishi}(1996)}]{DBW2:Takaishi:1996xj}
\bibinfo{author}{\bibfnamefont{T.}~\bibnamefont{Takaishi}},
  \bibinfo{journal}{Phys. Rev.} \textbf{\bibinfo{volume}{D54}},
  \bibinfo{pages}{1050} (\bibinfo{year}{1996}).

\bibitem[{\citenamefont{de~Forcrand et~al.}(2000)}]{DBW2:deForcrand:1999bi}
\bibinfo{author}{\bibfnamefont{P.}~\bibnamefont{de~Forcrand}}
  \bibnamefont{et~al.} (\bibinfo{collaboration}{QCD-TARO}),
  \bibinfo{journal}{Nucl. Phys.} \textbf{\bibinfo{volume}{B577}},
  \bibinfo{pages}{263} (\bibinfo{year}{2000}),
  \eprint[http://arXiv.org/abs]{hep-lat/9911033}.

\bibitem[{\citenamefont{Orginos}(2002)}]{DW-DBW2:Orginos:2001xa}
\bibinfo{author}{\bibfnamefont{K.}~\bibnamefont{Orginos}}
  (\bibinfo{collaboration}{RBC}), \bibinfo{journal}{Nucl. Phys. Proc. Suppl.}
  \textbf{\bibinfo{volume}{106}}, \bibinfo{pages}{721} (\bibinfo{year}{2002}),
  \eprint[http://arXiv.org/abs]{hep-lat/0110074}.

\bibitem[{\citenamefont{Hasenfratz and Alexandru}(2002)}]{Hasenfratz:2002ym}
\bibinfo{author}{\bibfnamefont{A.}~\bibnamefont{Hasenfratz}} \bibnamefont{and}
  \bibinfo{author}{\bibfnamefont{A.}~\bibnamefont{Alexandru}},
  \bibinfo{journal}{Phys. Rev.} \textbf{\bibinfo{volume}{D65}},
  \bibinfo{pages}{114506} (\bibinfo{year}{2002}),
  \eprint[http://arXiv.org/abs]{hep-lat/0203026}.

\bibitem[{\citenamefont{Alexandru and Hasenfratz}(2002)}]{Alexandru:2002jr}
\bibinfo{author}{\bibfnamefont{A.}~\bibnamefont{Alexandru}} \bibnamefont{and}
  \bibinfo{author}{\bibfnamefont{A.}~\bibnamefont{Hasenfratz}}
  (\bibinfo{year}{2002}), \eprint[http://arXiv.org/abs]{hep-lat/0207014}.

\bibitem[{\citenamefont{Weisz}(1983)}]{Weisz:1983zw}
\bibinfo{author}{\bibfnamefont{P.}~\bibnamefont{Weisz}},
  \bibinfo{journal}{Nucl. Phys.} \textbf{\bibinfo{volume}{B212}},
  \bibinfo{pages}{1} (\bibinfo{year}{1983}).

\bibitem[{\citenamefont{Luscher and Weisz}(1985)}]{LW:Luscher:1985xn}
\bibinfo{author}{\bibfnamefont{M.}~\bibnamefont{Luscher}} \bibnamefont{and}
  \bibinfo{author}{\bibfnamefont{P.}~\bibnamefont{Weisz}},
  \bibinfo{journal}{Commun. Math. Phys.} \textbf{\bibinfo{volume}{97}},
  \bibinfo{pages}{59} (\bibinfo{year}{1985}).

\bibitem[{\citenamefont{Bliss et~al.}(1996)\citenamefont{Bliss, Hornbostel, and
  Lepage}}]{LW-tadpole:Bliss:1996wy}
\bibinfo{author}{\bibfnamefont{D.~W.} \bibnamefont{Bliss}},
  \bibinfo{author}{\bibfnamefont{K.}~\bibnamefont{Hornbostel}},
  \bibnamefont{and} \bibinfo{author}{\bibfnamefont{G.~P.} \bibnamefont{Lepage}}
   (\bibinfo{year}{1996}), \eprint[http://arXiv.org/abs]{hep-lat/9605041}.

\bibitem[{\citenamefont{Bernard and DeGrand}(2000)}]{Bernard:1999kc}
\bibinfo{author}{\bibfnamefont{C.~W.} \bibnamefont{Bernard}} \bibnamefont{and}
  \bibinfo{author}{\bibfnamefont{T.}~\bibnamefont{DeGrand}},
  \bibinfo{journal}{Nucl. Phys. Proc. Suppl.} \textbf{\bibinfo{volume}{83}},
  \bibinfo{pages}{845} (\bibinfo{year}{2000}),
  \eprint[http://arXiv.org/abs]{hep-lat/9909083}.

\bibitem[{\citenamefont{Bernard et~al.}(2002)}]{Bernard:2002pc}
\bibinfo{author}{\bibfnamefont{C.}~\bibnamefont{Bernard}} \bibnamefont{et~al.}
  (\bibinfo{collaboration}{MILC})  (\bibinfo{year}{2002}),
  \eprint[http://arXiv.org/abs]{hep-lat/0206016}.

\bibitem[{\citenamefont{Albanese et~al.}(1987)}]{APE:Albanese:1987ds}
\bibinfo{author}{\bibfnamefont{M.}~\bibnamefont{Albanese}} \bibnamefont{et~al.}
  (\bibinfo{collaboration}{APE}), \bibinfo{journal}{Phys. Lett.}
  \textbf{\bibinfo{volume}{B192}}, \bibinfo{pages}{163} (\bibinfo{year}{1987}).

\bibitem[{\citenamefont{Lepage and Mackenzie}(1993)}]{Lepage:1993xa}
\bibinfo{author}{\bibfnamefont{G.~P.} \bibnamefont{Lepage}} \bibnamefont{and}
  \bibinfo{author}{\bibfnamefont{P.~B.} \bibnamefont{Mackenzie}},
  \bibinfo{journal}{Phys. Rev.} \textbf{\bibinfo{volume}{D48}},
  \bibinfo{pages}{2250} (\bibinfo{year}{1993}),
  \eprint[http://arXiv.org/abs]{hep-lat/9209022}.

\bibitem[{\citenamefont{Lee and Sharpe}(2002)}]{LeeSharpe:Lee:2002bf}
\bibinfo{author}{\bibfnamefont{W.-j.} \bibnamefont{Lee}} \bibnamefont{and}
  \bibinfo{author}{\bibfnamefont{S.~R.} \bibnamefont{Sharpe}}
  (\bibinfo{year}{2002}), \eprint[http://arXiv.org/abs]{hep-lat/0208036}.

\bibitem[{\citenamefont{DeGrand}(2002)}]{DeGrand:2002va}
\bibinfo{author}{\bibfnamefont{T.}~\bibnamefont{DeGrand}}
  (\bibinfo{year}{2002}), \eprint[http://arXiv.org/abs]{hep-lat/0210028}.

\bibitem[{\citenamefont{DeGrand}(1999)}]{DeGrand:1999gp}
\bibinfo{author}{\bibfnamefont{T.}~\bibnamefont{DeGrand}}
  (\bibinfo{collaboration}{MILC}), \bibinfo{journal}{Phys. Rev.}
  \textbf{\bibinfo{volume}{D60}}, \bibinfo{pages}{094501}
  (\bibinfo{year}{1999}), \eprint[http://arXiv.org/abs]{hep-lat/9903006}.

\bibitem[{\citenamefont{Gupta et~al.}(1997)\citenamefont{Gupta, Bhattacharya,
  and Sharpe}}]{Gupta:1997yt}
\bibinfo{author}{\bibfnamefont{R.}~\bibnamefont{Gupta}},
  \bibinfo{author}{\bibfnamefont{T.}~\bibnamefont{Bhattacharya}},
  \bibnamefont{and} \bibinfo{author}{\bibfnamefont{S.~R.}
  \bibnamefont{Sharpe}}, \bibinfo{journal}{Phys. Rev.}
  \textbf{\bibinfo{volume}{D55}}, \bibinfo{pages}{4036} (\bibinfo{year}{1997}),
  \eprint[http://arXiv.org/abs]{hep-lat/9611023}.

\bibitem[{\citenamefont{Edwards et~al.}(1998)\citenamefont{Edwards, Heller, and
  Klassen}}]{Potential:Edwards:1998xf}
\bibinfo{author}{\bibfnamefont{R.~G.} \bibnamefont{Edwards}},
  \bibinfo{author}{\bibfnamefont{U.~M.} \bibnamefont{Heller}},
  \bibnamefont{and} \bibinfo{author}{\bibfnamefont{T.~R.}
  \bibnamefont{Klassen}}, \bibinfo{journal}{Nucl. Phys.}
  \textbf{\bibinfo{volume}{B517}}, \bibinfo{pages}{377} (\bibinfo{year}{1998}),
  \eprint[http://arXiv.org/abs]{hep-lat/9711003}.

\bibitem[{\citenamefont{Hasenfratz et~al.}(2001)\citenamefont{Hasenfratz,
  Hoffmann, and Knechtli}}]{Hasenfratz:2001tw}
\bibinfo{author}{\bibfnamefont{A.}~\bibnamefont{Hasenfratz}},
  \bibinfo{author}{\bibfnamefont{R.}~\bibnamefont{Hoffmann}}, \bibnamefont{and}
  \bibinfo{author}{\bibfnamefont{F.}~\bibnamefont{Knechtli}}
  (\bibinfo{year}{2001}), \eprint[http://arXiv.org/abs]{hep-lat/0110168}.

\bibitem[{\citenamefont{Gattringer et~al.}(2001)\citenamefont{Gattringer,
  Hoffmann, and Schaefer}}]{Gattringer:2001jf}
\bibinfo{author}{\bibfnamefont{C.}~\bibnamefont{Gattringer}},
  \bibinfo{author}{\bibfnamefont{R.}~\bibnamefont{Hoffmann}}, \bibnamefont{and}
  \bibinfo{author}{\bibfnamefont{S.}~\bibnamefont{Schaefer}}
  (\bibinfo{year}{2001}), \eprint[http://arXiv.org/abs]{hep-lat/0112024,
  accepted for publication in Phys. Rev. D}.

\bibitem[{\citenamefont{Necco}(2002)}]{Necco:2002zy}
\bibinfo{author}{\bibfnamefont{S.}~\bibnamefont{Necco}}
  (\bibinfo{year}{2002}), \eprint[http://arXiv.org/abs]{hep-lat/0208052}.

\bibitem[{\citenamefont{DeGrand et~al.}(1997)\citenamefont{DeGrand, Hasenfratz,
  and Kovacs}}]{DeGrand:1997gu}
\bibinfo{author}{\bibfnamefont{T.}~\bibnamefont{DeGrand}},
  \bibinfo{author}{\bibfnamefont{A.}~\bibnamefont{Hasenfratz}},
  \bibnamefont{and} \bibinfo{author}{\bibfnamefont{T.~G.}
  \bibnamefont{Kovacs}}, \bibinfo{journal}{Nucl. Phys.}
  \textbf{\bibinfo{volume}{B505}}, \bibinfo{pages}{417} (\bibinfo{year}{1997}),
  \eprint[http://arXiv.org/abs]{hep-lat/9705009}.

\bibitem[{\citenamefont{Hasenfratz and Nieter}(1998)}]{Hasenfratz:1998qk}
\bibinfo{author}{\bibfnamefont{A.}~\bibnamefont{Hasenfratz}} \bibnamefont{and}
  \bibinfo{author}{\bibfnamefont{C.}~\bibnamefont{Nieter}},
  \bibinfo{journal}{Phys. Lett.} \textbf{\bibinfo{volume}{B439}},
  \bibinfo{pages}{366} (\bibinfo{year}{1998}),
  \eprint[http://arXiv.org/abs]{hep-lat/9806026}.

\bibitem[{\citenamefont{Hasenfratz}(2001)}]{Hasenfratz:2001wd}
\bibinfo{author}{\bibfnamefont{A.}~\bibnamefont{Hasenfratz}},
  \bibinfo{journal}{Phys. Rev.} \textbf{\bibinfo{volume}{D64}},
  \bibinfo{pages}{074503} (\bibinfo{year}{2001}),
  \eprint[http://arXiv.org/abs]{hep-lat/0104015}.

\bibitem[{\citenamefont{Schafer and Shuryak}(1998)}]{Schafer:1998wv}
\bibinfo{author}{\bibfnamefont{T.}~\bibnamefont{Schafer}} \bibnamefont{and}
  \bibinfo{author}{\bibfnamefont{E.~V.} \bibnamefont{Shuryak}},
  \bibinfo{journal}{Rev. Mod. Phys.} \textbf{\bibinfo{volume}{70}},
  \bibinfo{pages}{323} (\bibinfo{year}{1998}), \eprint{hep-ph/9610451}.

\bibitem[{\citenamefont{Hernandez et~al.}(1999)\citenamefont{Hernandez, Jansen,
  and Luscher}}]{Hernandez:1998et}
\bibinfo{author}{\bibfnamefont{P.}~\bibnamefont{Hernandez}},
  \bibinfo{author}{\bibfnamefont{K.}~\bibnamefont{Jansen}}, \bibnamefont{and}
  \bibinfo{author}{\bibfnamefont{M.}~\bibnamefont{Luscher}},
  \bibinfo{journal}{Nucl. Phys.} \textbf{\bibinfo{volume}{B552}},
  \bibinfo{pages}{363} (\bibinfo{year}{1999}),
  \eprint[http://arXiv.org/abs]{hep-lat/9808010}.

\bibitem[{\citenamefont{Kovacs}(2002)}]{Kovacs:2002nz}
\bibinfo{author}{\bibfnamefont{T.~G.} \bibnamefont{Kovacs}}
  (\bibinfo{year}{2002}), \eprint[http://arXiv.org/abs]{hep-lat/0209125}.

\end{thebibliography}

\end{document}